\documentclass[a4paper,fleqn]{cas-sc}

\usepackage[numbers]{natbib}
\usepackage{caption}
\usepackage{subcaption}

\def\tsc#1{\csdef{#1}{\textsc{\lowercase{#1}}\xspace}}
\tsc{WGM}
\tsc{QE}
\begin{document}
\let\WriteBookmarks\relax
\def\floatpagepagefraction{1}
\def\textpagefraction{.001}

\shorttitle{Development and test of a mini-Data Acquisition System}    

\shortauthors{Yuxiang Guo et al.}  

\title [mode = title]{Development and test of a mini-Data Acquisition system for the High-Luminosity LHC upgrade of the ATLAS Monitored Drift Tube detector}  

%% Title footnote mark
%% eg: \tnotemark[1]
%\tnotemark[<tnote number>] 

%% Title footnote 1.
%% eg: \tnotetext[1]{Title footnote text}
%\tnotetext[<tnote number>]{<tnote text>} 


%\author{Yuxiang Guo}
%\fnmark[a]

%% Corresponding author indication
%\cormark[1]
\author[1]{Yuxiang Guo}[orcid=0000-0002-6027-5132]
\ead{gyuxiang@umich.edu}

\credit{<Credit authorship details>}

\affiliation[1]{organization={Department of Physics, University of Michigan},
	% addressline={450 Church St}, 
	city={Ann Arbor},
	state={MI},
	%          citysep={}, % Uncomment if no comma needed between city and postcode
	postcode={48109},
	country={USA}}
	
\author[1]{Xueye Hu}[orcid=0000-0002-3617-290X]

\author[1]{Thomas Schwarz}[orcid=0000-0001-5660-2690]

\author[1]{Bing Zhou}[orcid=0000-0002-0034-6576]

\author[1]{Junjie Zhu}[orcid=0000-0002-5278-2855]
\ead{junjie@umich.edu}


\cortext[1]{Corresponding author}


\begin{abstract}
	New front-end electronics including ASICs and FPGA boards are under development for the ATLAS Monitored Drift Tube (MDT) detector to handle the large data rates and harsh environment expected at high-luminosity LHC runs. 
	A mobile Data Acquisition (miniDAQ) system is designed to perform integration tests of these front-end electronics. In addition, it will be used for surface commissioning of 96 small-radius MDT (sMDT) chambers and for integration and commissioning of new front-end electronics on the present ATLAS MDT chambers. Details of the miniDAQ hardware and firmware are described in this article. The miniDAQ system is also used to read out new front-end electronics on an sMDT prototype chamber using cosmic muons and results obtained are shown. 
\end{abstract}


%% Research highlights
%\item 
%\item 
%\item 

% Keywords

\maketitle

% Main text
\section{Introduction}\label{sec-intro}
The ATLAS detector~\cite{ATLAS:2008xda} is a general-purpose particle detector designed to study proton-proton interactions at the Large Hadron Collider (LHC) at CERN~\cite{Evans:2008zzb}. Its muon spectrometer is designed to trigger and identify muons produced in collisions and to measure their momenta~\cite{ATLAS:2010xrj}. Resistive Plate Chambers (RPC)~\cite{ATLAS:2021pft} and Thin Gap Chambers (TGC)~\cite{Nagai:1996mf} are used as fast trigger chambers in the barrel ($|\eta| < 1.05$) \footnote{ATLAS uses a right-handed coordinate system with its origin at the nominal interaction point (IP) in the center of the detector
and the $z$-axis along the beam pipe. The $x$-axis points from the IP to the center of the LHC ring, and the $y$-axis points upwards. 
Cylindrical coordinates ($r$, $\phi$) are used in the transverse plane. The pseudorapidity is defined in terms of the polar angle $\theta$ as $\eta= -\ln\tan(\theta/2)$.} and endcap ($1.05 < |\eta| < 2.4$) regions, respectively. Three (two) stations of Monitored Drift Tube (MDT) chambers in the barrel (endcap) regions are used as precision-tracking chambers. The endcap inner station has a new detector composed of eight layers of Micromegas (MM) and eight layers of small-strip TGC (sTGC) installed in 2021~\cite{Kawamoto:2013udg}. 

The ATLAS muon spectrometer is designed to provide a standalone momentum measurement with a relative resolution better than 3\% over a wide $p_T$ range and 10\% at $p_T=1$ TeV~\cite{ATLAS:2008xda}, where $p_T$ is the muon transverse momentum. The momentum in the muon spectrometer is measured from the deflection of the muon trajectory in the magnetic field generated by a system of air-core toroid coils. The MDT chambers, together with MM and sTGC, perform precision charged particle tracking in the transverse plane. Each MDT tube has an average spatial resolution of $\sim 100$ $\mu$m.  

The current MDT readout system was designed to cope with the original LHC design luminosity of $10^{34}$ cm$^{-2}$ s$^{-1}$ with a first-level trigger acceptance rate of 100 kHz and a latency of 2.5 $\mu$s~\cite{Arai:2008zzb}. Hits are stored in buffered memories of front-end electronics waiting for the ATLAS first-level trigger acceptance signal, while the first-level muon trigger is exclusively based on RPC/TGC trigger detectors. %with excellent time resolution ($1-25$ ns) but modest spatial resolution ($\sim 2$ cm). 
The high-luminosity LHC (HL-LHC) will have the instantaneous luminosity increased by a factor of $5 - 7.5$ and the integrated luminosity increased by a factor of $\sim 10$. The targeted ATLAS first-level trigger acceptance rate and latency are 1 MHz and 10 $\mu$s, respectively~\cite{ATLAS:2017tdaq}. 
However, each MDT tube has a diameter of 3 cm and ionized electrons near the tube wall can take up to 750 ns to reach the anode wire, far longer than the LHC bunch crossing time interval of 25 ns. This long drift time makes the current MDT readout scheme inappropriate for future HL-LHC runs since a hit is likely to be read out multiple times at 1 MHz. A triggerless mode to send all muon hits off chambers to new trigger and readout circuitry is preferred. In addition, the relatively long trigger latency allows ATLAS to include MDT chamber data at the first trigger level to improve the trigger muon momentum resolution and to reject low momentum muons~\cite{Richter:2020tvi}. 

In addition to upgrading all MDT front-end and back-end electronics, ATLAS also plans to replace part of the barrel inner station MDT chambers with small-radius MDT (sMDT) chambers. These MDT chambers will have the tube radius reduced by a factor of 2, to gain space to add an additional layer of RPC chambers in that station to increase the overall RPC detector coverage~\cite{ATLAS:2017muon}. In total 96 new sMDT chambers will be constructed and installed later~\cite{Eberwein:2021nvt}. 

Extensive developments are ongoing for the MDT front-end and back-end electronics for the HL-LHC upgrade. A mobile data acquisition (miniDAQ) system is needed to integrate various front-end electronics prototypes. This system will also be critical for surface commissioning of these 96 sMDT chambers and for integration and commissioning of new front-end electronics on the present MDT chambers inside the ATLAS collision hall.

\section{The miniDAQ system}
\subsection{Introduction of the MDT front-end and back-end electronics}
Each MDT tube has a diameter of 3 cm and a wall thickness of 400 $\mu$m. The central tungsten wire has a diameter of 50 $\mu$m. Tubes are operated with a gas mixture of Ar/CO$_2$ (93\%/7\%) at 3 bar absolute pressure. For each track, the electrons from primary ionization clusters drift to the central wire along radial lines. The induced signal propagates along the wire where it is read out by the MDT front-end electronics. The difference between the earliest arrival time of the signal at the wire and the reference time provided by trigger chambers gives the drift time of the muon hit, and this drift time is used to determine the drift radius. 

The signal from each tube is first processed by a custom-designed Amplifier/Shaper/Discriminator (ASD) ASIC~\cite{Kroha:2016fid, DeMatteis:2017xky}. A discriminator is used to determine the signal arrival time, the time when the signal crosses a predefined threshold. This time depends on the signal pulse height which results in a degradation of the time resolution. The resolution degradation can partially be recovered by applying a time skew correction using the integrated charge of the signal pulse. A Wilkinson analog-to-digital converter (ADC) is introduced inside the ASD to integrate the signal pulse over a predefined integration window of $\sim 20$ ns. The total collected charge is measured by the discharge time of a capacitor by a rundown current. The signal arrival time (also called the leading edge time) and the discharge time (also called the trailing edge time) are converted into ADC counts using a time-to-digital converter (TDC) ASIC with a bin size of 0.78 ns~\cite{Wang:2017jnd, Liang:2019weg, Guo:2020zyb}. 
To avoid multiple hits from multiple threshold crossings of a single signal, the ASD ASIC can be programmed with a dead time of $\sim 1~\mu$s. After the detection of the earliest arrival signal, there are no additional time measurements performed within this dead time. 

Each ASD ASIC can handle 8 tubes and each TDC can handle discriminated signals from three ASDs. A mezzanine card with three ASDs and one TDC thus handles 24 tubes. A Chamber Service Module (CSM) multiplexes data from up to 18 mezzanine cards and sends these data via an optical module (VTRx+)~\cite{Soos:2017stv} to the MDT Trigger Processor~\cite{Cieri:2020bfv}, where the relevant hits are extracted out of the raw data stream. Pattern recognition, segment-finding, and track-fitting algorithms are then applied to determine the muon momentum at the first trigger level. Hit data are stored for transmission to a network called Front End LInk eXchange (FELIX)~\cite{Wu:2018rnc} after receiving the first-level trigger acceptance signal.

\subsection{Introduction of the miniDAQ system}
Due to the new MDT trigger and readout scheme, all front-end and back-end electronics need to be redesigned. Both ASD and TDC designs have been finished. All ASD chips have been produced, while all TDC chips are expected to be produced in 2022. The designs for both mezzanine cards and CSM are close to be final and minor modifications to the current prototypes are expected. The MDT Data Processor is still under development. 

It is critical to design a miniDAQ system to integrate these prototype ASICs and boards together and to demonstrate that the new front-end electronics can run in the triggerless mode to send out all hits. The miniDAQ system is a lightweight version of the MDT Data Processor. It will send out all matched hits to a PC for storage and pattern recognition, segment-finding, and track-fitting algorithms will be performed offline. As a result, a low-cost FPGA can be used and a FELIX system is not needed. The miniDAQ system is expected to be mobile and can be used to study the performance of new sMDT chambers. It is also expected to be used for the integration and commissioning of new front-end electronics on the present MDT chambers inside the ATLAS collision hall. 

\begin{figure}[h]
	\centering
	\includegraphics[width=0.6\textwidth]{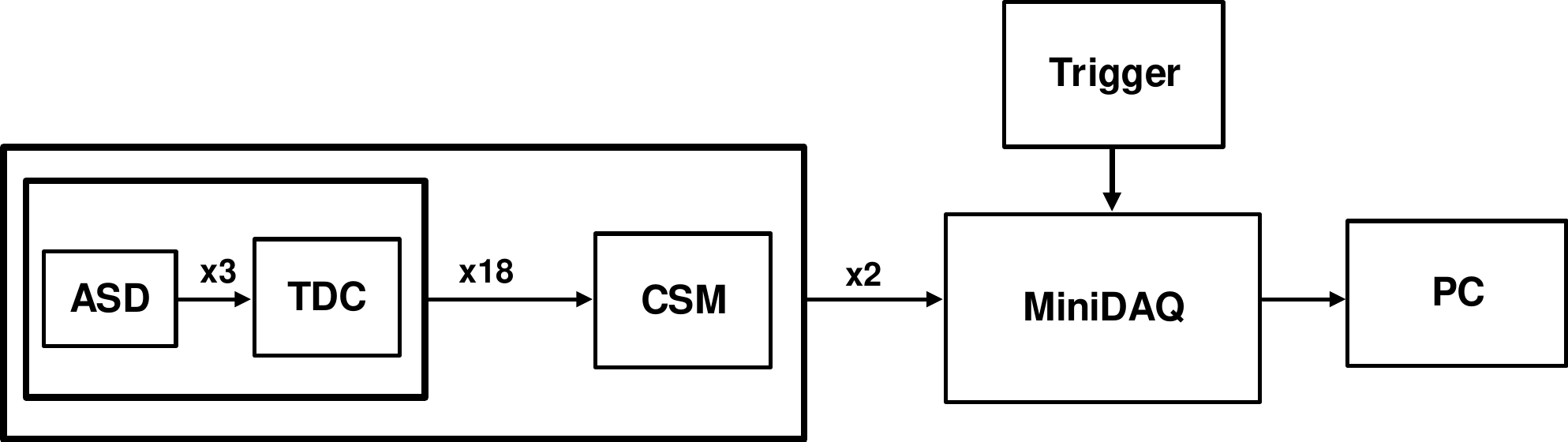}
	\caption{The miniDAQ readout system used to read out data from two CSMs. Each CSM will handle up to 18 mezzanine cards, and each mezzanine card has three ASD ASICs and one TDC ASIC. Trigger signals can be either the external coincidence trigger signals or programmed trigger logic signals. The output data will be sent to a PC for storage.}
	\label{fig:MDTDAQ_MiniDAQ}
\end{figure}

Figure~\ref{fig:MDTDAQ_MiniDAQ} shows the overall miniDAQ readout system planned for a single (s)MDT chamber. Due to the smaller tube radius, an sMDT chamber can have more than 500 tubes, thus two CSMs are needed to read out all tubes. The miniDAQ system is designed to handle at least 108 ASD and 36 TDC ASICs on 36 mezzanine cards. All TDCs can be configured to run in either the triggerless or the trigger mode, and the default mode is the triggerless mode~\cite{Liang:2019weg}. The miniDAQ system also receives data from at least two CSMs. Matched hits with the arrival time within the trigger time window are sent to a PC using an ethernet port.

The requirements of the miniDAQ system are the following: 1) configure all ASICs and boards connected; 2) distribute Trigger, Timing, and Control (TTC) signals to each CSM via a downlink fiber at a line rate of 2.56 Gbps; 3) receive the CSM output optical signals, and each CSM has two serial optical uplink fibers running at a line rate of 10.24 Gbps; 4) receive the trigger signal (can be either the external coincidence trigger signal or a programmed trigger logic signal) and perform trigger matching to only send matched hit data for offline storage; (5) monitor voltages and temperatures of all mezzanine cards connected; and 6) monitor detector data in real time.  

\subsection{The miniDAQ hardware}
\begin{figure}[h]
	\centering
	\includegraphics[width=0.75\textwidth]{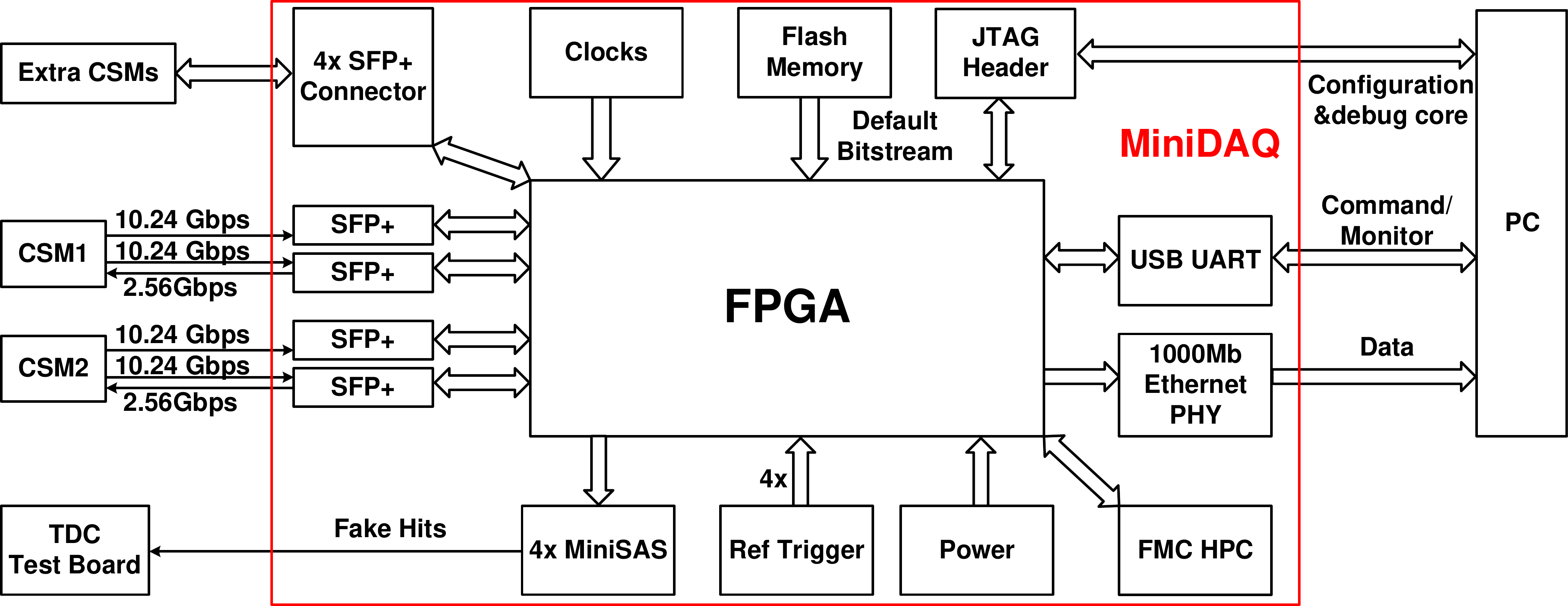}
	\caption{Connections between the front-end boards, the miniDAQ board, and a PC. The red box indicates major components on the miniDAQ board. The central part of the board is a Xilinx FPGA.}
	\label{fig:MiniDAQ_diagram}
\end{figure}

\begin{figure}[h]
	
	\centering
	\includegraphics[width=0.5\textwidth]{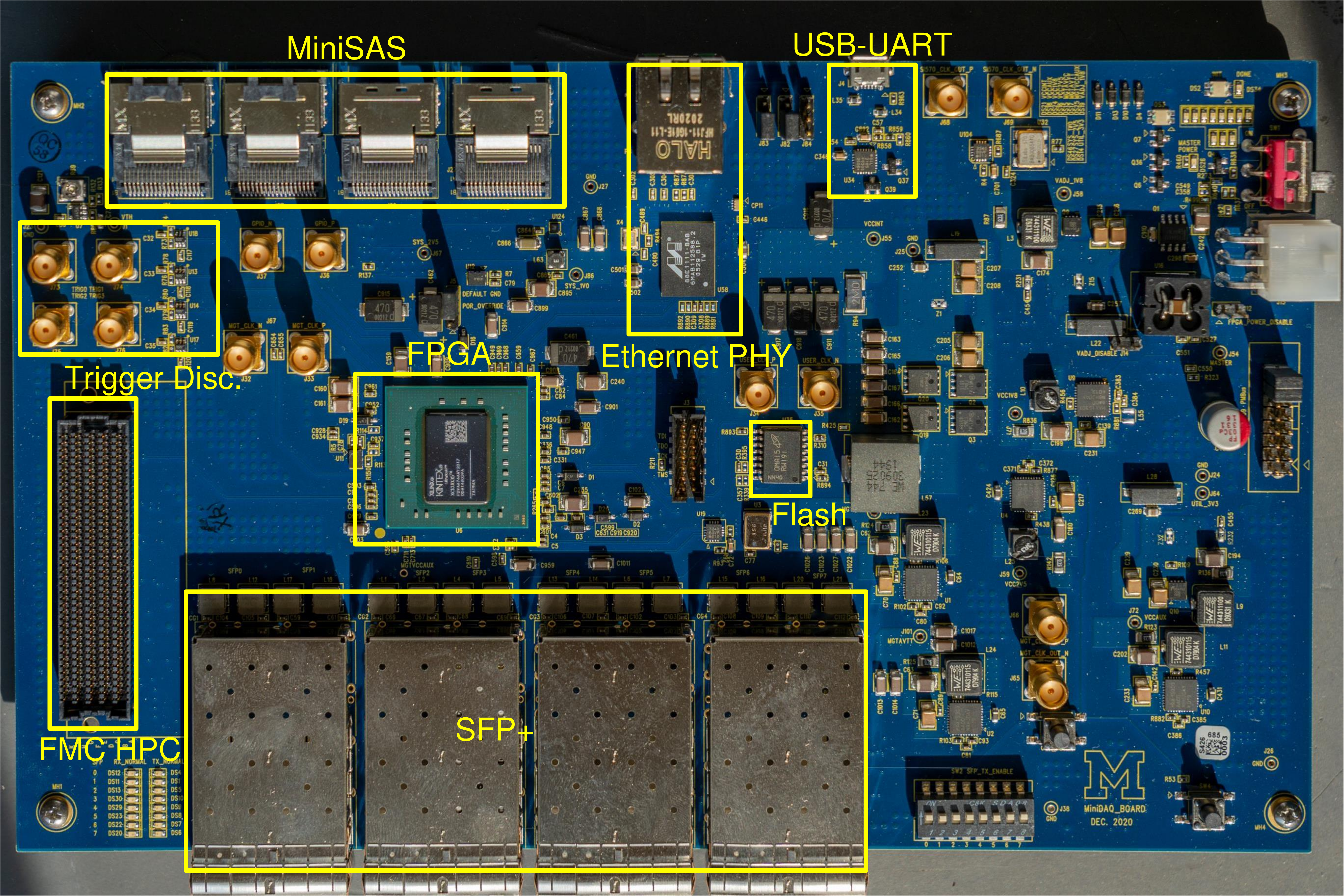}
	\caption{Picture of an miniDAQ board. Major components and connectors are indicated.}
	\label{fig:MiniDAQ_board}
\end{figure}

Figure~\ref{fig:MiniDAQ_diagram} shows the connections of the miniDAQ system. A miniDAQ board is designed to communicate between front-end electronics and a PC. Figure~\ref{fig:MiniDAQ_board} shows a picture of the actual miniDAQ board. 

The central part of the miniDAQ board is a Xilinx Ultrascale KU035-1FBVA676C FPGA. It contains about 444k logic cells, 19 Mb Block RAM, and 16 GTH transceivers with a maximum line rate of 12.5 Gbps each. In addition, the FPGA has 10 Clock Management tiles, 104 high-range I/Os, and 208 high-performance I/Os. 

Since each CSM uses two uplink fibers and one downlink fiber, in total four SFP+ modules are needed to handle two CSMs. Four additional SFP+ modules are reserved to handle two additional CSMs if needed. 
A 400 I/O FPGA Mezzanine Card (FMC) High-pin count (HPC) connector is placed on the lower left corner of the board and is connected to four FPGA GTH transceivers. This connector provides flexibility for extra devices (such as an additional SFP+ adapter board to integrate two extra CSMs).

\begin{figure}[h]
	\centering
	\includegraphics[width=0.45\textwidth]{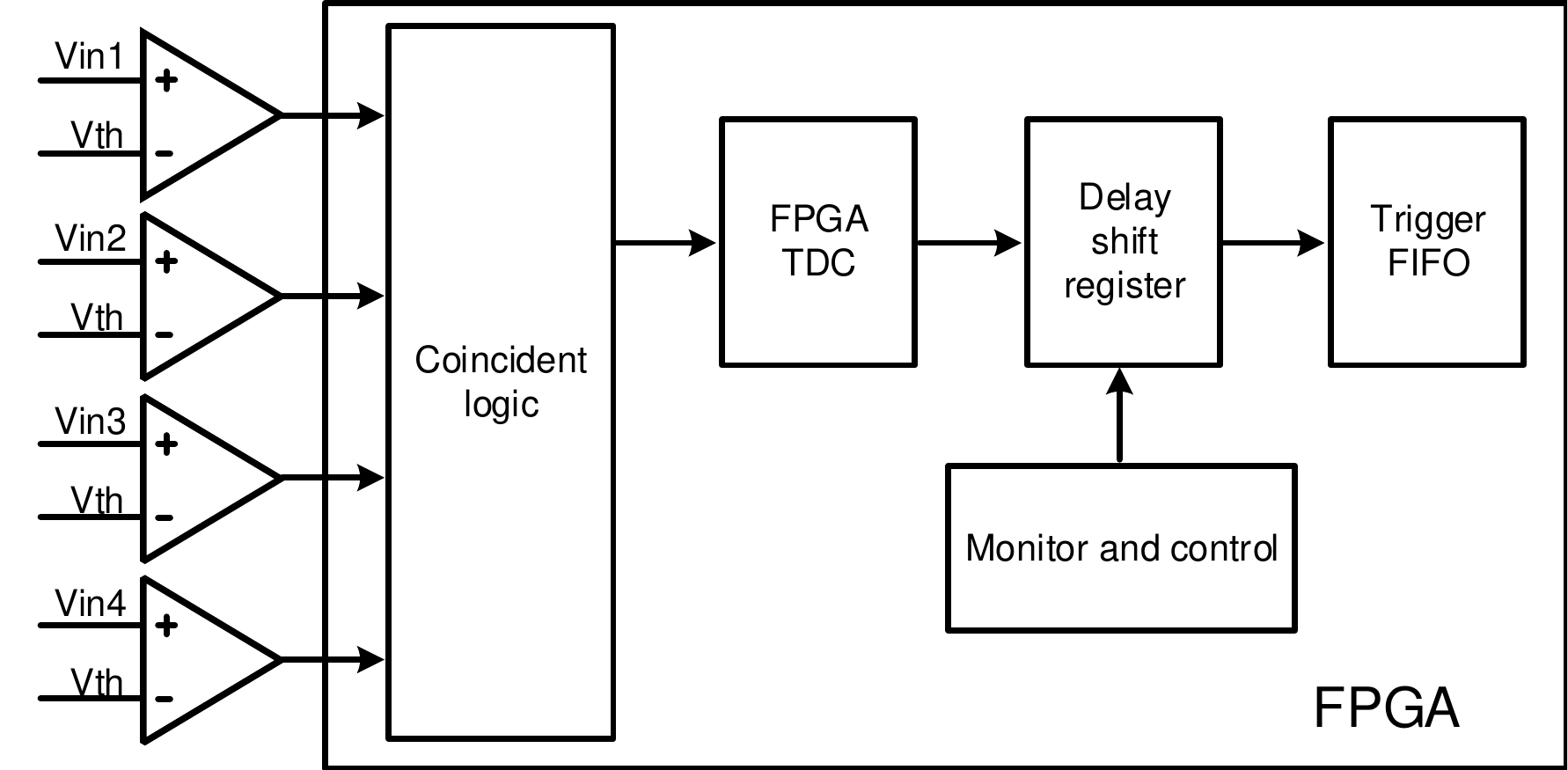}
	\caption{Trigger option of the miniDAQ board. An adjustable threshold is applied to all comparators. Threshold-crossing time after the coincidence logic is digitized by the FPGA-TDC and then used in the trigger matching process after a programmable delay.}
	\label{fig:MiniDAQ_trigger}
\end{figure}

Four SMA connectors are used to receive trigger signals, as shown in Fig.~\ref{fig:MiniDAQ_trigger}. Each channel contains a high-speed comparator (ADCMP604) and the comparator outputs are connected to the on-board FPGA. The trigger signal can be either the analog signals from photomultipliers (PMTs) connnected to plastic scintillators or the final coincidence signal of these analog signals. When multiple PMTs are used, the coincidence can be made inside the FPGA firmware. The coincidence signal is then digitized with a TDC implemented in the FPGA firmware (FPGA-TDC). After a programmable delay, the digitized signal queues in the first-in first-out buffer (FIFO) to start the trigger matching process.

The miniDAQ board features an on-board 320 MHz oscillator, providing the reference clock for the GTH transceivers. External clock can also be fed into for synchronization if multiple miniDAQ boards are used. A flash memory is used to boot the firmware into the FPGA when the board is powered on. 

The board exchanges commands and front-end electronics monitoring information with a PC via a USB-UART interface, and the detector data is sent out for offline storage through the gigabit ethernet interface. Four miniSAS connectors are reserved for joint test with a TDC test board. DC-DC converters and low-dropout regulators are used to provide power to the FPGA and all other on-board devices.

\subsection{The miniDAQ firmware}
\begin{figure}[h]
	\centering
	\includegraphics[width=0.8\textwidth]{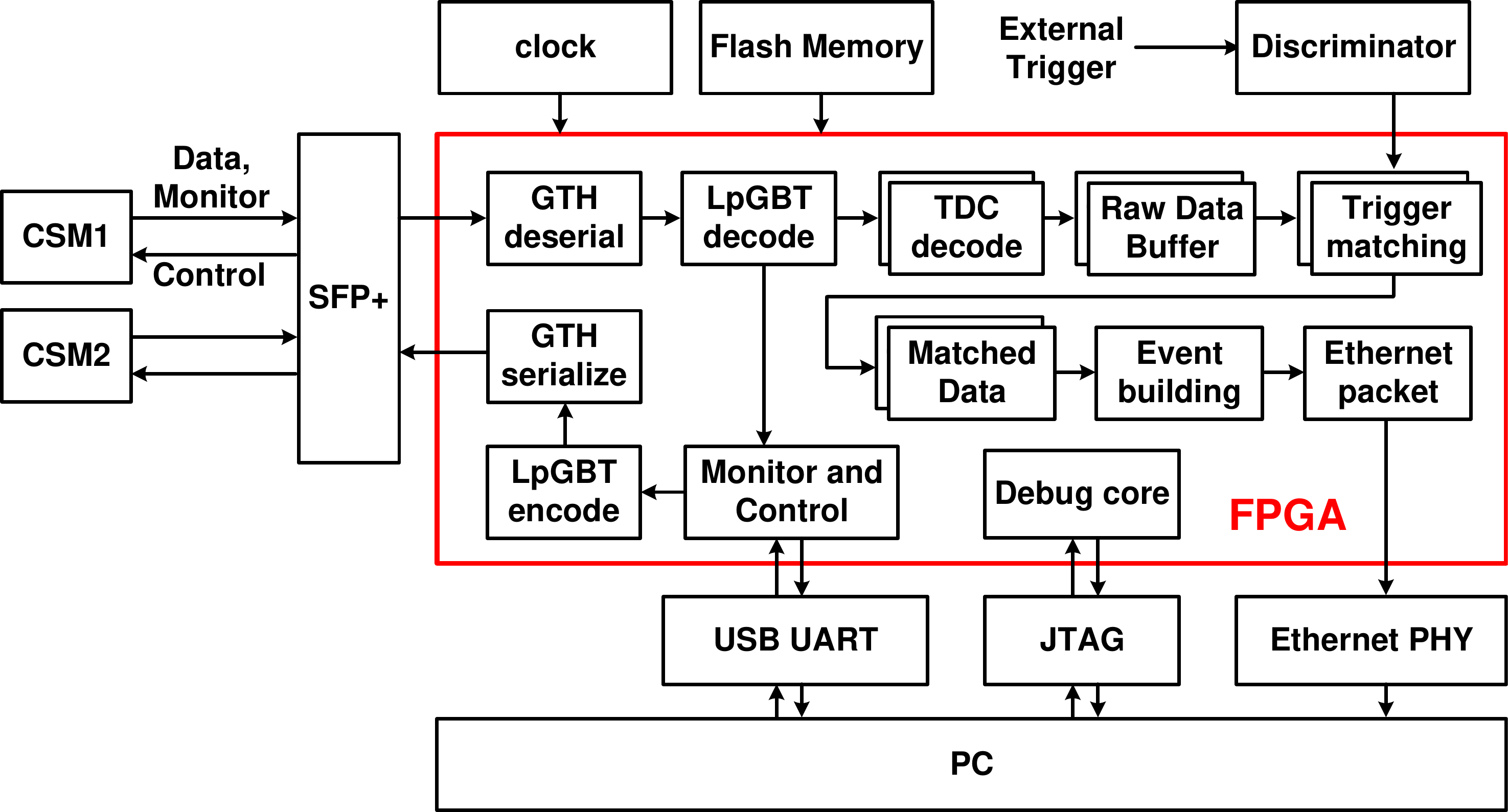}
	\caption{Functional block diagram of the miniDAQ system. Main FPGA firmware modules are shown in the red box and the arrows indicate data-flow directions.}
	\label{fig:MiniDAQ_function_diagram}
\end{figure}

Figure~\ref{fig:MiniDAQ_function_diagram} shows the functional block diagram of the miniDAQ system. The FPGA firmware can handle uplink and downlink data of the CSMs, digitize the external trigger from scintillators, and communicate with a PC for offline data storage and command exchange. The uplink data is passed through the multi-stage decoding, the trigger matching, and sent out in packets to a PC via the ethernet interface. The front-end configuration commands are exchanged with the PC, then encoded and serialized as the downlink data. A JTAG debug core is also used for online debugging.

The key to successfully store the detector timing data for offline analysis is to reduce the overall data bandwidth and select events of interest. When two CSMs are connected to the miniDAQ system (shown in Fig.~\ref{fig:MDTDAQ_MiniDAQ}), the line rate of the uplink data adds up to 40.96 Gbps. While this rate suits the maximum data rate requirement of the HL-LHC upgrade, data filtering and trigger matching are a must for the offline data storage as either background noises or idle words occupy more than 90\% of the uplink bandwidth. The miniDAQ firmware runs data decoding and trigger matching in parallel for each uplink. 

\begin{figure}[h]
	\centering
	\includegraphics[width=0.8\textwidth]{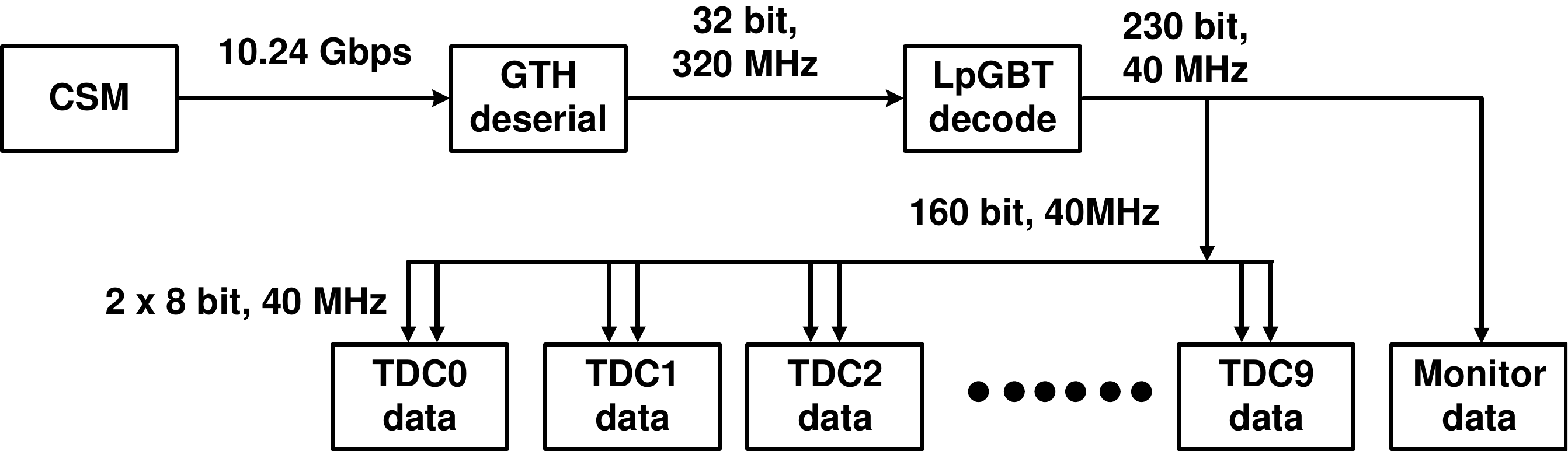}
	\caption{Decoding scheme of the miniDAQ system for one uplink. Serial data from one uplink goes through different de-serial and decoding stages to form data for each individual TDC and data for voltage/temperature monitoring.}
	\label{fig:MiniDAQ_decode_logic_en}
\end{figure}

Multi-stage data decoding for one uplink is processed as illustrated in Fig.~\ref{fig:MiniDAQ_decode_logic_en}. The 10.24 Gbps serial data is firstly sampled and de-serialized by the Xilinx GTH data interface, yielding a 32-bit word at 320 MHz. Secondly, the lpGBT-FPGA IP further de-interleaves, decodes, and de-scrambles the uplink data to a 230-bit word at 40 MHz according to the GBT protocol. This 230-bit word contains 160-bit TDC data from 10 mezzanine cards and 70-bit data from voltage and temperature sensors. The TDC data are then grouped into 10 slots corresponding to its original mezzanine number and are decoded individually. 

The MDT TDC utilizes two e-links running at 320 Mbps for its 8b/10b-encoded output data, one for even bits and one for odd bits. When there are no hits to be sent out, the TDC generates idle words continuously, which are useful for data alignment when decoding at the receiver end. Even and odd bits are aligned, concatenated and decoded by the miniDAQ system. At this stage, idle words are thrown out and hits are buffered in on-chip random-access memories (RAMs). While the throughput data rate is reduced by removing idle words, a trigger matching process is implemented to select interesting hits from the background noise.

The arrival time of a trigger signal is digitized by the FPGA-TDC. The FPGA-TDC utilizes multiple phases of the reference clock, with the help of the FPGA built-in clock manager modules to achieve the same time bin size as the front-end TDC ASIC. After a pre-set latency during when all expected hits are being collected in the RAMs, the trigger matching process compares the leading edge time information between the trigger signal and the hit data stored in the RAMs. Hits within a fixed time window relative to the trigger signal are packed with the trigger data as an valid event and sent out through the gigabit ethernet. An internal counter monitors the hits in the RAM and rejects the outdated ones, with a programmable rejection window set based on the trigger latency and trigger matching window.

\begin{figure}[h]
	\centering
	\includegraphics[width=0.55\textwidth]{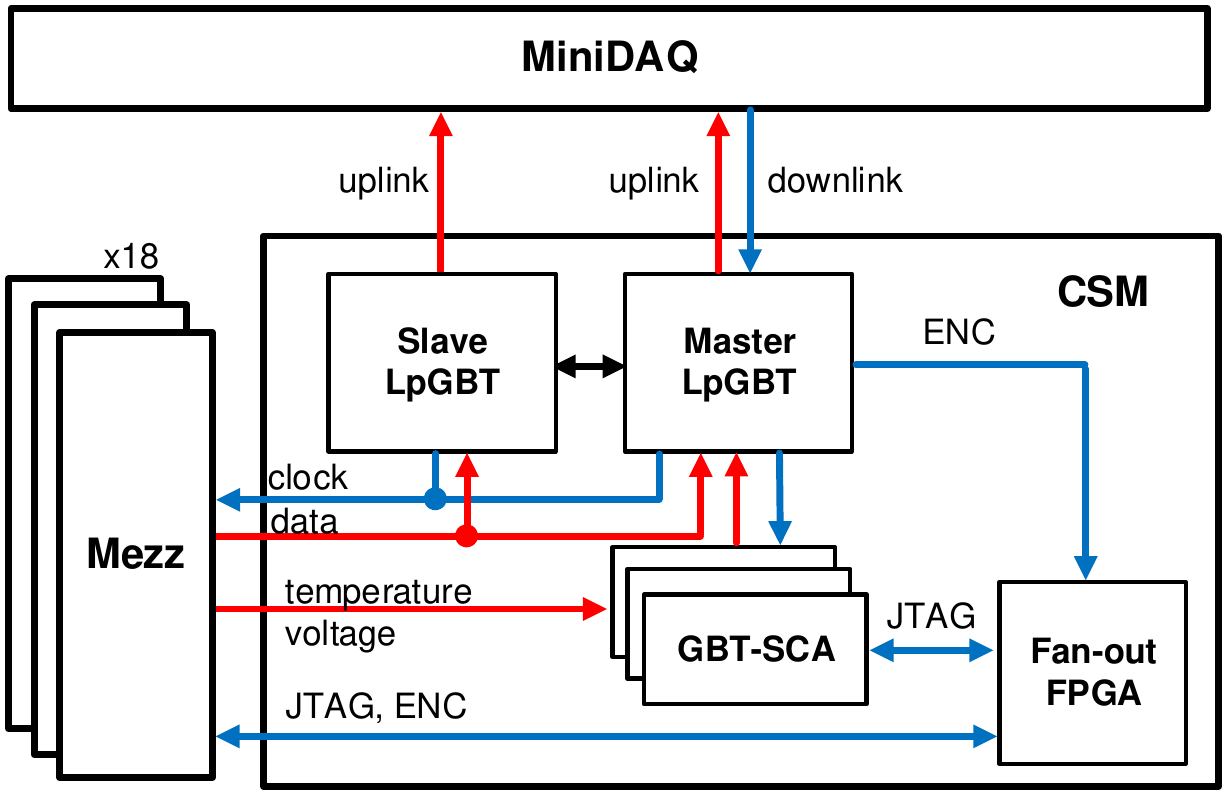}
	\caption{The uplink (red) and downlink (blue) data flow of the front-end electronics. Chamber data, as well as voltage/temperature monitoring data, are sent by uplinks. Configuration and TTC signals are sent by downlinks. }
	\label{fig:front_control}
\end{figure}

The miniDAQ system sends front-end ASIC configuration bits and the encoded control (ENC) signals for bunch count reset and system reset through the downlink data, as indicated in Fig.~\ref{fig:front_control}. The two lpGBT ASICs~\cite{lpGBT2}, which work in the master/slave mode, are configured directly through their serial control interfaces. The configuration of the lpGBT ASICs enables the clock distribution and the data sampling of the mazzanine cards.

The configuration of the mezzanine cards is achieved by utilizing the JTAG master in the GBT-SCA ASIC~\cite{GBT-SCA}, which converts the serial downlink data into JTAG signals and provides a configurable clock rate. The converted JTAG signals in the GBT-SCA, along with the ENC signals directly from the downlink, are distributed to all mezzanine cards through the fan-out FPGA on the CSM board.

\section{Results from cosmic-ray studies}
\begin{figure}[h]
	\centering
	\includegraphics[width=0.48\textwidth]{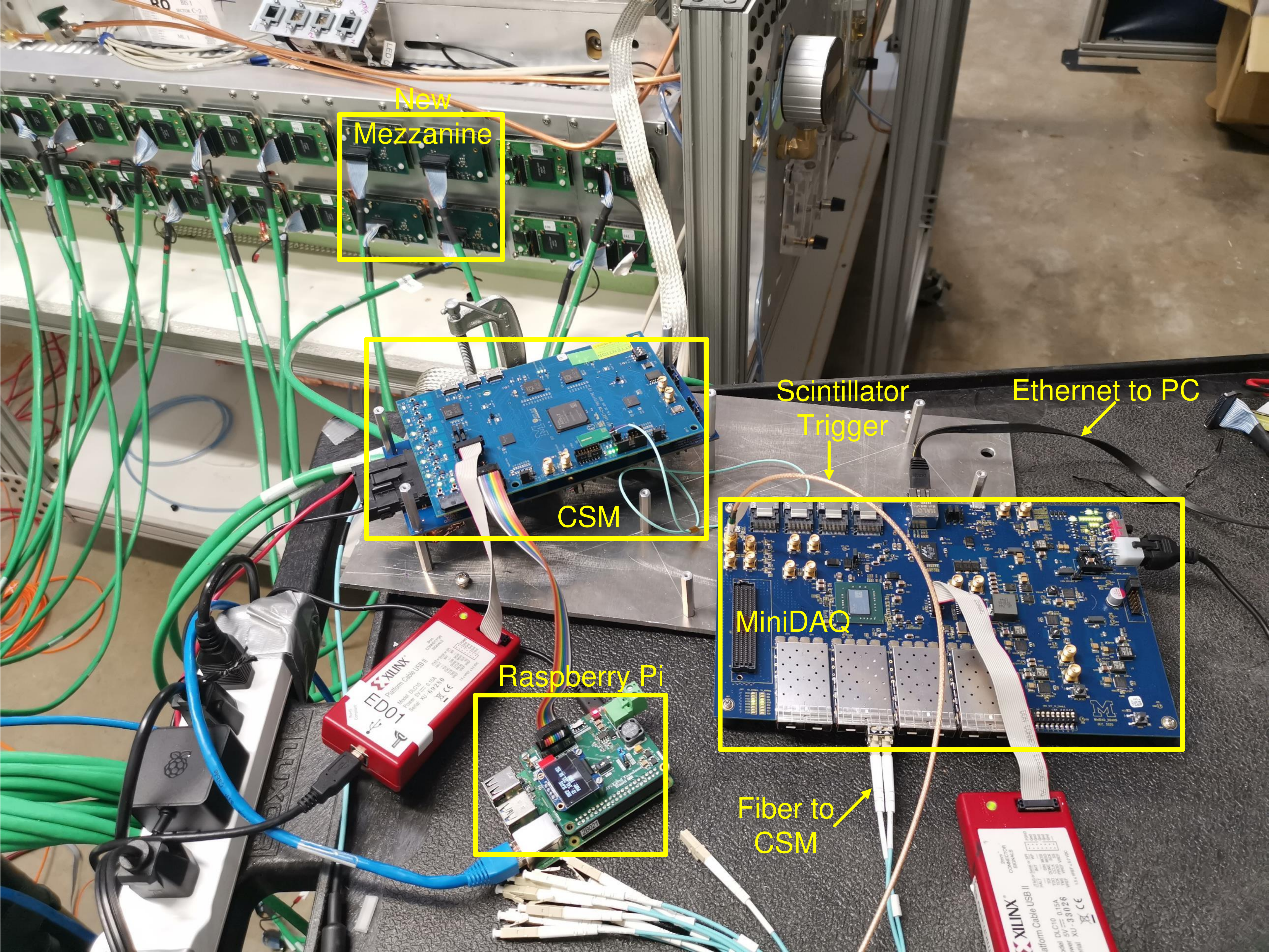}
	\caption{Setup of the cosmic-ray test station with an sMDT chamber. Detector signals are processed and digitized by ASD and TDC ASICs on the mezzanine card. All digitized hit data are sent to the CSM and then to the miniDAQ board. The miniDAQ board also receives the coincidence signal from two PMTs connected to a large area plastic scintillator, and only matched hits within the allowed time window are sent out to a PC via an Ethernet port. A Raspberry Pi board is used to configure two lpGBT ASICs on the CSM.}
	\label{fig:minidaq_onchamber}
\end{figure}

The miniDAQ system is used to read out an sMDT prototype chamber built at the University of Michigan. 
Figure~\ref{fig:minidaq_onchamber} illustrates the setup of the detector and the miniDAQ system. The sMDT prototype chamber has 
eight layers of drift tubes with 70 tubes per layer and a tube diameter of 1.5 cm. Tubes are filled with a gas mixture
of 93\% Argon and 7\% CO$_2$ at 3-bar absolute pressure. A high voltage of 2,730 volts is applied on the central wire. 
Due to the availability of prototype boards, 
four three-layer stacked mezzanine card designed by the ATLAS group at Max Planck Institute for Physics at Munich are used to read out 96 tubes. 
Each set of stacked card has three ASD ASICs and one TDCv2 ASIC mounted. A new CSM board is used to collect output data from these four TDCs. 
A Raspberry Pi board is also used to configure two lpGBT ASICs into ready status on the CSM board. The lpGBT output is converted to optical signal and sent by VTRx+ to the miniDAQ system via optical fibers. 
Trigger and reference time information are generated by a 1 m $\times$ 2 m scintillator placed on the top of the chamber. 
The signals from two PMTs placed at both ends of the scintillator are discriminated and a coincidence trigger signal is formed. 
The resulting trigger signal is sent to the miniDAQ board and all matched hits are transmitted to a PC via the Ethernet port. 

\begin{figure}[h]
	\centering
	\includegraphics[width=0.8\textwidth]{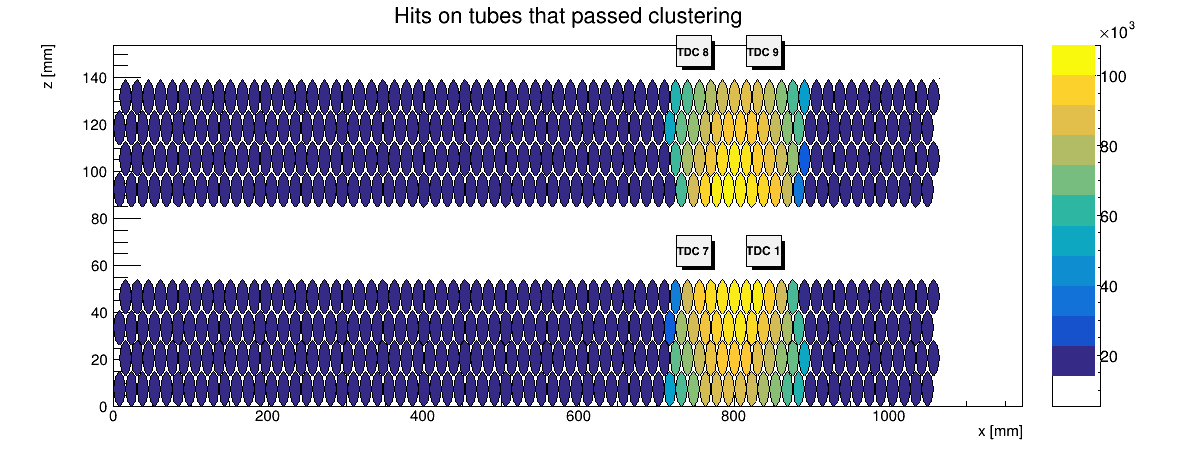}
	\caption{Occupancy of tubes that are read out. Only 96 tubes are read out by the new electronics. A muon track with at least 6 tubes being hit is required for each event.}
	\label{fig:preselected_hits}
\end{figure}

Cosmic-ray data has been taken to study the performance of both the detector and front-end electronics.
Figure~\ref{fig:preselected_hits} shows the occupancy of the tubes that are read out by the new electronics. In total 96 tubes 
(48 on the top multilayer and 48 on the bottom multilayer) are connected to the four mezzanine cards. A muon track with at least six tubes being hit is required to be reconstructed in each event. That explains low occupancy for tubes around the edges since most muon tracks passing through them are dropped during the reconstruction.

\begin{figure}[ht!]
	\centering
	% \hfill
	\begin{subfigure}[b]{0.45\textwidth}
		\centering
		\includegraphics[width=\textwidth,height=\linewidth,keepaspectratio]{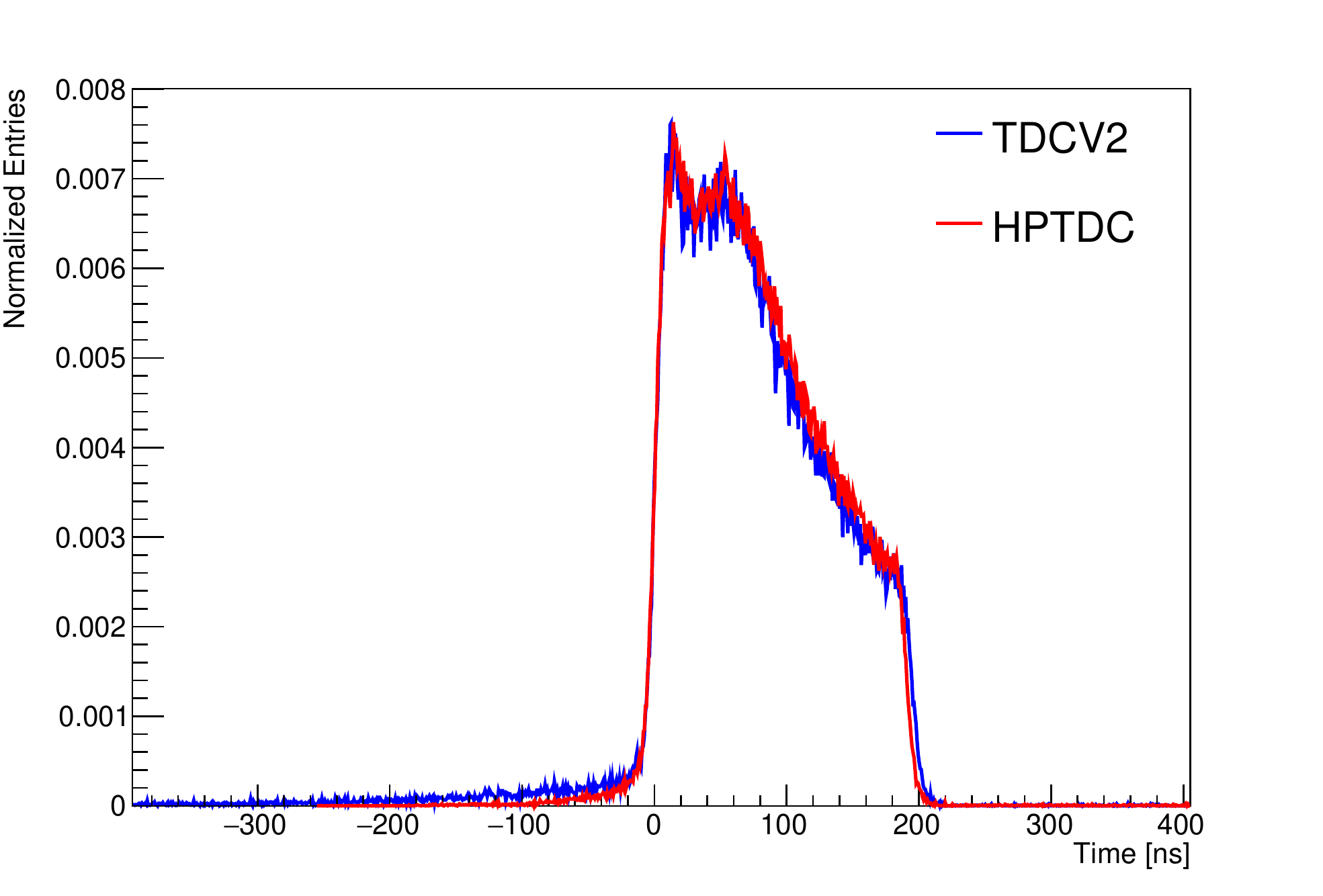}
		\caption{}
	\end{subfigure}
	% \hfill
	\begin{subfigure}[b]{0.45\textwidth}
		\centering
		\includegraphics[width=\textwidth,height=\linewidth,keepaspectratio]{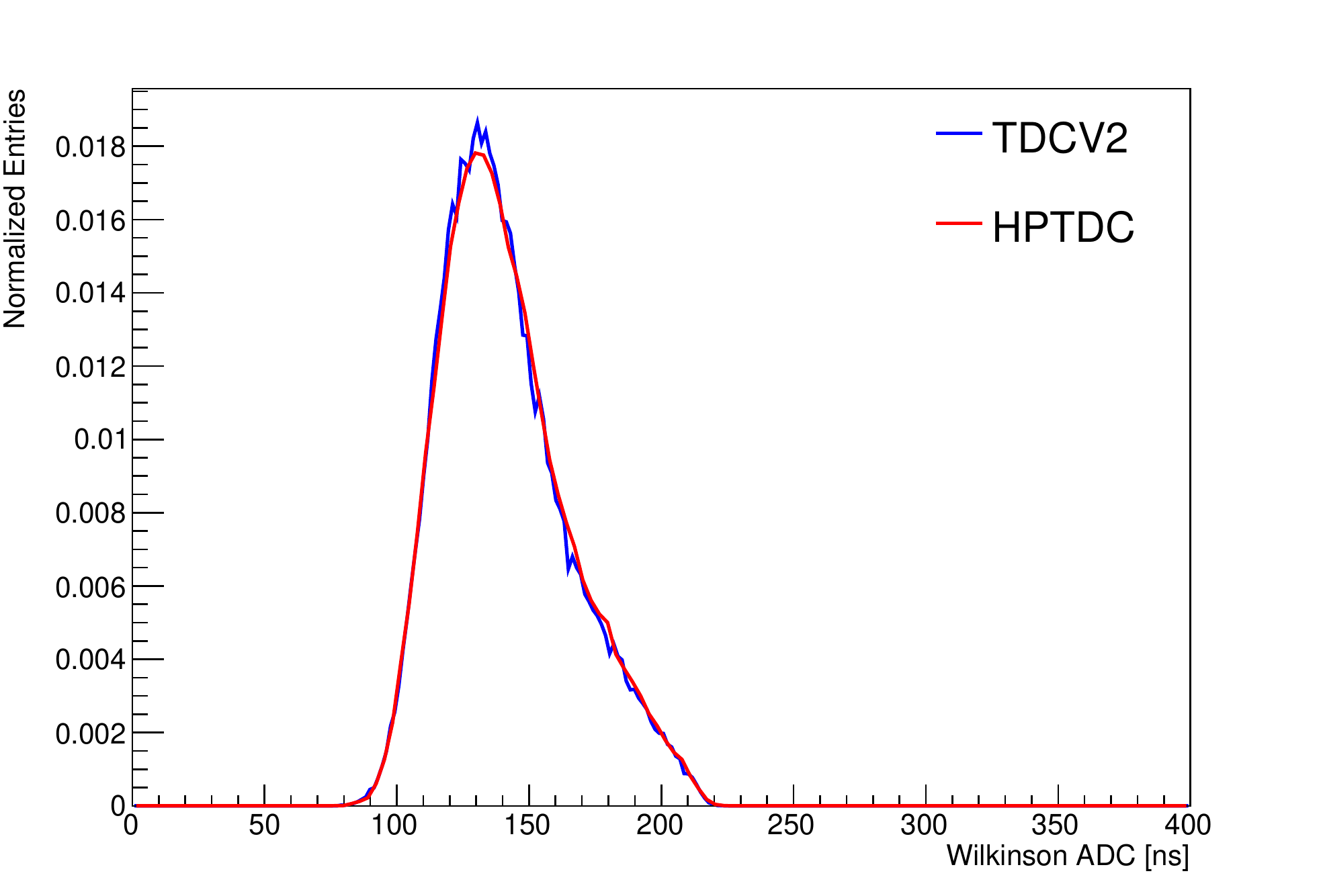}
		\caption{}
	\end{subfigure}	
	\caption{Comparisons between the TDCv2 prototype and the CERN-designed HPTDC ASIC for (a) the signal arrival time spectrum, and (b) the pulse height spectrum for a single channel. Cosmic muon data are used.}
	\label{fig:spec_compare}%
\end{figure}

%When a muon passes through the region read out by the four cards, it will ionize gas atoms and 
%electrons from the primary ionization clusters drift to the central wire due to the electric field. 
%The earliest arrival signal at the wire is captured by the ASD ASIC. 
%The difference between the earliest arrival time of the signal at the wire and the reference time provided by the scintillator gives the drift time of the muon hit. 
%In addition, ASD measures the pulse height of the detector signal for the first $\sim 20$ ns. 
%The pulse height is encoded as the time interval between the leading and trailing edges
%of the ASD output pulse. TDC digitizes the output pulses from three ASDs with a time precision of 780 ps. 
Figure~\ref{fig:spec_compare} shows the measured signal arrival time and pulse height spectra for cosmic-ray muons for a single tube. The width 
of the signal arrival time spectrum is $\sim 200$ ns, which corresponds to the time needed for ionized electrons produced close to the tube wall to drift to the central wire. 
The spectra measured with the TDCv2 prototype ASIC are also compared with the results obtained using a mezzanine board with a CERN-designed HPTDC ASIC~\cite{HPTDC_paper}. Good agreement is observed.

Detailed studies are performed on the data collected to obtain the spatial resolution and detection efficiency. 
The drift-distance-to-drift-time relation (also called the $r(t)$ function) is measured and used to obtain the drift radius for each fired tube. 
A straight line is then used to fit these draft radii and both biased and unbiased tracking residuals are obtained. The residual is the difference between the drift radius
and the radius predicted by the straight line fit. Biased residuals are those from fits using all tube
hits, whereas unbiased residuals are found by refitting a track multiple times by removing one hit
from the fit and finding the residual of the hit removed from the fit.
More details about the track reconstruction can be found in Ref.~\cite{Nelson:2021xxm}. 

\begin{figure}[ht!]
	\centering
	% \hfill
	\begin{subfigure}[b]{0.45\textwidth}
		\centering
		\includegraphics[width=\textwidth,height=\linewidth,keepaspectratio]{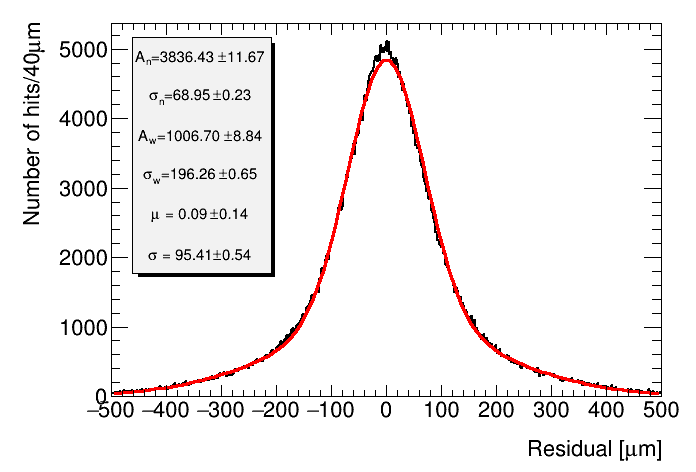}
		\caption{}
	\end{subfigure}
	% \hfill
	\begin{subfigure}[b]{0.45\textwidth}
		\centering
		\includegraphics[width=\textwidth,height=\linewidth,keepaspectratio]{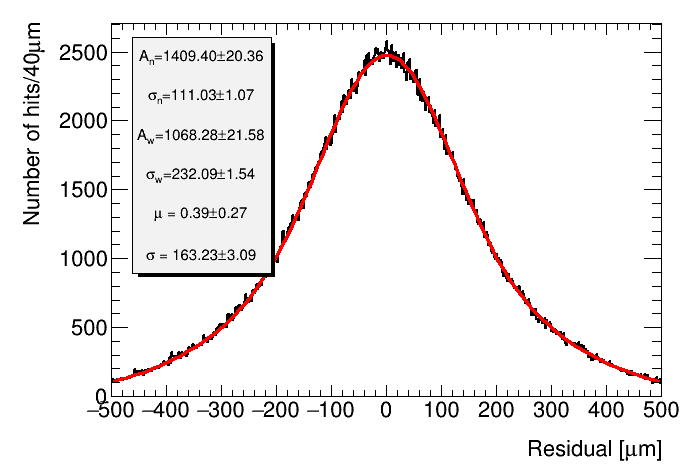}
		\caption{}
	\end{subfigure}	
	\caption{Residual distribution of the cosmic muon data for (a) biased residual, and (b) unbiased residual. The red lines indicate the results from the double Gaussian fit.}%
	\label{fig:residual_fit}%
\end{figure}

\begin{figure}[ht!]
	\centering
	% \hfill
	\begin{subfigure}[b]{0.4\textwidth}
		\centering
		\includegraphics[width=\textwidth,height=\linewidth,keepaspectratio]{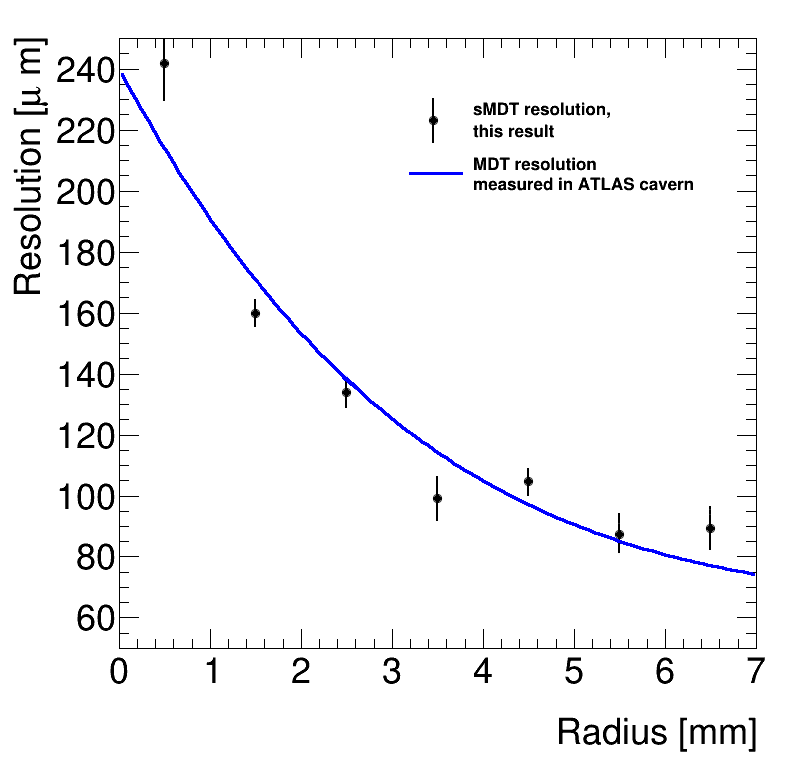}
		\caption{}
	\end{subfigure}
	% \hfill
	\begin{subfigure}[b]{0.4\textwidth}
		\centering
		\includegraphics[width=\textwidth,height=\linewidth,keepaspectratio]{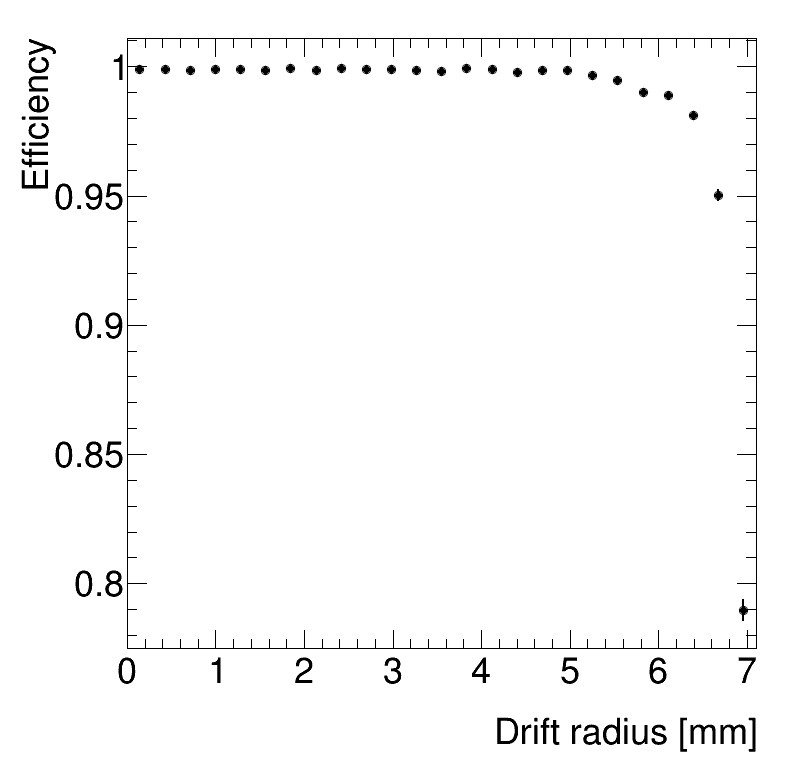}
		\caption{}
	\end{subfigure}	
	\caption{Results from cosmic muon studies for (a) spatial resolution as a function of the drift radius (the solid line shows the resolution measured for the current ATLAS MDT detector installed inside the collision hall), and (b) detection efficiency as a function of the drift radius.}
	\label{fig:efficiency}%
\end{figure}

Figure~\ref{fig:residual_fit} shows the biased and unbiased residual distributions. Two Gaussian distributions with the same mean value are used to fit the 
residual distributions. The average width of the fit ($\sigma_b$ for the biased residual and $\sigma_u$ for the unbiased residual) is defined as the two fitted Gaussian widths weighted by the amplitude of each Gaussian and is found to be $\sigma_b=95.41 \pm 0.54$ $\mu$m and $\sigma_b=163.23 \pm 3.09$ $\mu$m. The spatial resolution $\sigma$, defined as $\sqrt{\sigma_b \sigma_u}$, is $124.8 \pm 1.2$ $\mu$m. It has to be noted that the numbers presented are before subtracting the contribution of the multiple scattering. Due to low energies of cosmic muons detected, multiple scattering effects need to be subtracted. A detailed Monte Carlo simulation is performed \cite{Nelson:2021xxm} and the spatial resolution as a function of the drift radius with multiple scattering effects removed is shown in Fig.~\ref{fig:efficiency}(a). The best fit curve observed in the Run 2 MDT resolution study~\cite{ATLAS:2019mcj} is also shown for comparison purpose. 

The muon detection efficiency as a function of the drift radius is shown in Fig.~\ref{fig:efficiency}(b). To evaluate whether a track passes through a tube volume, 
that tube is excluded from the track fit to not bias the fit to be closer to the tube. Hits greater than $5~\sigma$ away from a track are not counted towards the efficiency~\cite{Nelson:2021xxm}.
The region near the tube wall is responsible for most of the inefficiency of the chamber. 

\section{Conclusions}
We present the design of a miniDAQ system for the HL-LHC upgrade of the ATLAS MDT detector. This miniDAQ system can configure all front-end ASIC and board prototypes, distribute TTC signals, and collect the CSM output data. It can also receive external trigger signals and perform trigger matching to select matched MDT hits for offline storage. Voltages and temperatures of all mezzanine cards are monitored, and detector data is monitored in real time. The miniDAQ system has been used successfully to read out 96 tubes from an sMDT prototype chamber using cosmic muons. Tube spatial resolution and detection efficiency are measured and are found to be consistent with results obtained from previous studies using the legacy ATLAS MDT electronics. This system is expected to be used for surface commissioning of 96 sMDT chambers to be built for the ATLAS HL-LHC upgrade. It will also be critical for future commissioning of new front-end electronics on the present MDT chambers inside the ATLAS collision hall.  

\section{Acknowledgement}
The work is supported by the US National Science Foundation (NSF) and the US Department of Energy (DOE) under contracts PHY1948993 (NSF) and DE-SC007859 (DOE).

\bibliographystyle{model1-num-names}
%\bibliographystyle{cas-model2-names}

\bibliography{refs}

\begin{thebibliography}{24}
\expandafter\ifx\csname natexlab\endcsname\relax\def\natexlab#1{#1}\fi
\providecommand{\url}[1]{\texttt{#1}}
\providecommand{\href}[2]{#2}
\providecommand{\path}[1]{#1}
\providecommand{\DOIprefix}{doi:}
\providecommand{\ArXivprefix}{arXiv:}
\providecommand{\URLprefix}{URL: }
\providecommand{\Pubmedprefix}{pmid:}
\providecommand{\doi}[1]{\href{http://dx.doi.org/#1}{\path{#1}}}
\providecommand{\Pubmed}[1]{\href{pmid:#1}{\path{#1}}}
\providecommand{\bibinfo}[2]{#2}
\ifx\xfnm\relax \def\xfnm[#1]{\unskip,\space#1}\fi
%Type = Article
\bibitem[{Aad et~al.(2008)}]{ATLAS:2008xda}
\bibinfo{author}{G.~Aad}, et~al. (\bibinfo{collaboration}{ATLAS}),
\newblock \bibinfo{title}{{The ATLAS Experiment at the CERN Large Hadron
  Collider}},
\newblock \bibinfo{journal}{JINST} \bibinfo{volume}{3} (\bibinfo{year}{2008})
  \bibinfo{pages}{S08003}.
%Type = Article
\bibitem[{Eva(2008)}]{Evans:2008zzb}
\bibinfo{title}{{LHC machine}},
\newblock \bibinfo{journal}{JINST} \bibinfo{volume}{3} (\bibinfo{year}{2008})
  \bibinfo{pages}{S08001}.
%Type = Article
\bibitem[{Aad et~al.(2010)}]{ATLAS:2010xrj}
\bibinfo{author}{G.~Aad}, et~al. (\bibinfo{collaboration}{ATLAS}),
\newblock \bibinfo{title}{{Commissioning of the ATLAS Muon Spectrometer with
  Cosmic Rays}},
\newblock \bibinfo{journal}{Eur. Phys. J. C} \bibinfo{volume}{70}
  (\bibinfo{year}{2010}) \bibinfo{pages}{875--916}.
%Type = Article
\bibitem[{Aad et~al.(2021)}]{ATLAS:2021pft}
\bibinfo{author}{G.~Aad}, et~al. (\bibinfo{collaboration}{ATLAS}),
\newblock \bibinfo{title}{{Performance of the ATLAS RPC detector and Level-1
  muon barrel trigger at $\sqrt{s}=13$ TeV}},
\newblock \bibinfo{journal}{JINST} \bibinfo{volume}{16} (\bibinfo{year}{2021})
  \bibinfo{pages}{P07029}.
%Type = Article
\bibitem[{Nagai(1996)}]{Nagai:1996mf}
\bibinfo{author}{K.~Nagai},
\newblock \bibinfo{title}{{Thin gap chambers in ATLAS}},
\newblock \bibinfo{journal}{Nucl. Instrum. Meth. A} \bibinfo{volume}{384}
  (\bibinfo{year}{1996}) \bibinfo{pages}{219--221}.
%Type = Article
\bibitem[{Kawamoto et~al.(2013)}]{Kawamoto:2013udg}
\bibinfo{author}{T.~Kawamoto}, et~al.,
\newblock \bibinfo{title}{{New Small Wheel Technical Design Report}}
  (\bibinfo{year}{2013}).
%Type = Article
\bibitem[{Arai et~al.(2008)}]{Arai:2008zzb}
\bibinfo{author}{Y.~Arai}, et~al.,
\newblock \bibinfo{title}{{ATLAS muon drift tube electronics}},
\newblock \bibinfo{journal}{JINST} \bibinfo{volume}{3} (\bibinfo{year}{2008})
  \bibinfo{pages}{P09001}.
%Type = Article
\bibitem[{Aad et~al.(2017)}]{ATLAS:2017tdaq}
\bibinfo{author}{G.~Aad}, et~al. (\bibinfo{collaboration}{ATLAS}),
\newblock \bibinfo{title}{{Technical Design Report for the Phase-II Upgrade of
  the ATLAS TDAQ System}}  (\bibinfo{year}{2017}).
%Type = Article
\bibitem[{Richter(2020)}]{Richter:2020tvi}
\bibinfo{author}{R.~Richter} (\bibinfo{collaboration}{ATLAS}),
\newblock \bibinfo{title}{{Upgrade of the ATLAS MDT Readout and Trigger for the
  HL-LHC}},
\newblock \bibinfo{journal}{PoS} \bibinfo{volume}{TWEPP2019}
  (\bibinfo{year}{2020}) \bibinfo{pages}{146}.
%Type = Article
\bibitem[{Aad et~al.(2017)}]{ATLAS:2017muon}
\bibinfo{author}{G.~Aad}, et~al. (\bibinfo{collaboration}{ATLAS}),
\newblock \bibinfo{title}{{Technical Design Report for the Phase-II Upgrade of
  the ATLAS muon spectrometer}}  (\bibinfo{year}{2017}).
%Type = Inproceedings
\bibitem[{Eberwein et~al.(2021)Eberwein, Kortner, Kroha, Rendel, Rieck, Soyk,
  Voevodina, and Walbrecht}]{Eberwein:2021nvt}
\bibinfo{author}{G.~H. Eberwein}, \bibinfo{author}{O.~Kortner},
  \bibinfo{author}{H.~Kroha}, \bibinfo{author}{M.~Rendel},
  \bibinfo{author}{P.~Rieck}, \bibinfo{author}{D.~Soyk},
  \bibinfo{author}{E.~Voevodina}, \bibinfo{author}{V.~Walbrecht}
  (\bibinfo{collaboration}{ATLAS Muon Group}),
\newblock \bibinfo{title}{{Commissioning and installation of the new
  small-Diameter Muon Drift Tube (sMDT) detectors for the Phase-I upgrade of
  the ATLAS Muon Spectrometer}},
\newblock in: \bibinfo{booktitle}{{2021 IEEE Nuclear Science Symposium (NSS)
  and Medical Imaging Conference (MIC) and 28th International Symposium on
  Room-Temperature Semiconductor Detectors}}, \bibinfo{year}{2021}.
  \href{http://arxiv.org/abs/2112.07026}{\tt arXiv:2112.07026}.
%Type = Inproceedings
\bibitem[{Kroha et~al.(2016)}]{Kroha:2016fid}
\bibinfo{author}{H.~Kroha}, et~al.,
\newblock \bibinfo{title}{{Performance of the new
  amplifier-shaper-discriminator chip for the ATLAS MDT chambers at the
  HL-LHC}},
\newblock in: \bibinfo{booktitle}{{2015 IEEE Nuclear Science Symposium and
  Medical Imaging Conference}}, \bibinfo{year}{2016}, p.
  \bibinfo{pages}{7581979}. \DOIprefix\doi{10.1109/NSSMIC.2015.7581979}.
  \href{http://arxiv.org/abs/1603.09093}{\tt arXiv:1603.09093}.
%Type = Article
\bibitem[{De~Matteis et~al.(2017)De~Matteis, Resta, Richter, Kroha, Fras, Zhao,
  Abovyan, and Baschirotto}]{DeMatteis:2017xky}
\bibinfo{author}{M.~De~Matteis}, \bibinfo{author}{F.~Resta},
  \bibinfo{author}{R.~Richter}, \bibinfo{author}{H.~Kroha},
  \bibinfo{author}{M.~Fras}, \bibinfo{author}{Y.~Zhao},
  \bibinfo{author}{S.~Abovyan}, \bibinfo{author}{A.~Baschirotto},
\newblock \bibinfo{title}{{An eight-channels 0.13-$\mu \text{m}$-CMOS front end
  for ATLAS muon-drift-tubes detectors}},
\newblock \bibinfo{journal}{IEEE Sensors J.} \bibinfo{volume}{17}
  (\bibinfo{year}{2017}) \bibinfo{pages}{3406--3415}.
%Type = Article
\bibitem[{Wang et~al.(2018)Wang, Liang, Xiao, An, Chapman, Dai, Zhou, Zhu, and
  Zhao}]{Wang:2017jnd}
\bibinfo{author}{J.~Wang}, \bibinfo{author}{Y.~Liang},
  \bibinfo{author}{X.~Xiao}, \bibinfo{author}{Q.~An}, \bibinfo{author}{J.~W.
  Chapman}, \bibinfo{author}{T.~Dai}, \bibinfo{author}{B.~Zhou},
  \bibinfo{author}{J.~Zhu}, \bibinfo{author}{L.~Zhao},
\newblock \bibinfo{title}{{Development of a time-to-digital converter ASIC for
  the upgrade of the ATLAS Monitored Drift Tube detector}},
\newblock \bibinfo{journal}{Nucl. Instrum. Meth. A} \bibinfo{volume}{880}
  (\bibinfo{year}{2018}) \bibinfo{pages}{174--180}.
%Type = Article
\bibitem[{Liang et~al.(2019)}]{Liang:2019weg}
\bibinfo{author}{Y.~Liang}, et~al.,
\newblock \bibinfo{title}{{Design and performance of a TDC ASIC for the upgrade
  of the ATLAS Monitored Drift Tube detector}},
\newblock \bibinfo{journal}{Nucl. Instrum. Meth. A} \bibinfo{volume}{939}
  (\bibinfo{year}{2019}) \bibinfo{pages}{10--15}.
%Type = Article
\bibitem[{Guo et~al.(2021)}]{Guo:2020zyb}
\bibinfo{author}{Y.~Guo}, et~al.,
\newblock \bibinfo{title}{{Design of a Time-to-Digital Converter ASIC and a
  mini-DAQ system for the Phase-2 upgrade of the ATLAS Monitored Drift Tube
  detector}},
\newblock \bibinfo{journal}{Nucl. Instrum. Meth. A} \bibinfo{volume}{988}
  (\bibinfo{year}{2021}) \bibinfo{pages}{164896}.
%Type = Article
\bibitem[{So\'os et~al.(2017)So\'os, D\'etraz, Olanter\"a, Sigaud, Troska,
  Vasey, and Zeiler}]{Soos:2017stv}
\bibinfo{author}{C.~So\'os}, \bibinfo{author}{S.~D\'etraz},
  \bibinfo{author}{L.~Olanter\"a}, \bibinfo{author}{C.~Sigaud},
  \bibinfo{author}{J.~Troska}, \bibinfo{author}{F.~Vasey},
  \bibinfo{author}{M.~Zeiler},
\newblock \bibinfo{title}{{Versatile Link PLUS transceiver development}},
\newblock \bibinfo{journal}{JINST} \bibinfo{volume}{12} (\bibinfo{year}{2017})
  \bibinfo{pages}{C03068}.
%Type = Article
\bibitem[{Cieri(2020)}]{Cieri:2020bfv}
\bibinfo{author}{D.~Cieri} (\bibinfo{collaboration}{ATLAS}),
\newblock \bibinfo{title}{{Hardware Demonstrator Of The Phase-II ATLAS MDT
  Trigger Processor}},
\newblock \bibinfo{journal}{PoS} \bibinfo{volume}{TWEPP2019}
  (\bibinfo{year}{2020}) \bibinfo{pages}{141}.
%Type = Article
\bibitem[{Wu(2019)}]{Wu:2018rnc}
\bibinfo{author}{W.~Wu} (\bibinfo{collaboration}{ATLAS TDAQ}),
\newblock \bibinfo{title}{{FELIX: the New Detector Interface for the ATLAS
  Experiment}},
\newblock \bibinfo{journal}{IEEE Trans. Nucl. Sci.} \bibinfo{volume}{66}
  (\bibinfo{year}{2019}) \bibinfo{pages}{986--992}.
%Type = Inproceedings
\bibitem[{Moreira(2020)}]{lpGBT2}
\bibinfo{author}{P.~Moreira},
\newblock \bibinfo{title}{{lpGBT project status and plans}},
\newblock in: \bibinfo{booktitle}{{ACES 2020 - Seventh Common ATLAS CMS
  Electronics Workshop for LHC Upgrades}}, \bibinfo{year}{2020}. \URLprefix
  \url{https://indico.cern.ch/event/863071/contributions/3738814/}.
%Type = Article
\bibitem[{Caratelli et~al.(2015)Caratelli, Bonacini, Kloukinas, Marchioro,
  Moreira, De~Oliveira, and Paillard}]{GBT-SCA}
\bibinfo{author}{A.~Caratelli}, \bibinfo{author}{S.~Bonacini},
  \bibinfo{author}{K.~Kloukinas}, \bibinfo{author}{A.~Marchioro},
  \bibinfo{author}{P.~Moreira}, \bibinfo{author}{R.~De~Oliveira},
  \bibinfo{author}{C.~Paillard},
\newblock \bibinfo{title}{{The GBT-SCA, a radiation tolerant ASIC for detector
  control and monitoring applications in HEP experiments}},
\newblock \bibinfo{journal}{JINST} \bibinfo{volume}{10} (\bibinfo{year}{2015})
  \bibinfo{pages}{C03034}.
%Type = Article
\bibitem[{Christiansen et~al.(2000)Christiansen, Marchioro, Moreira, Mota,
  Ryzhov, and Debieux}]{HPTDC_paper}
\bibinfo{author}{J.~Christiansen}, \bibinfo{author}{A.~Marchioro},
  \bibinfo{author}{P.~Moreira}, \bibinfo{author}{M.~Mota},
  \bibinfo{author}{V.~Ryzhov}, \bibinfo{author}{S.~Debieux},
\newblock \bibinfo{title}{{A data driven high performance time to digital
  converter}}  (\bibinfo{year}{2000}).
%Type = Article
\bibitem[{Nelson et~al.(2021)Nelson, Guo, Amidei, and Diehl}]{Nelson:2021xxm}
\bibinfo{author}{K.~Nelson}, \bibinfo{author}{Y.~Guo},
  \bibinfo{author}{D.~Amidei}, \bibinfo{author}{E.~Diehl},
\newblock \bibinfo{title}{{Performance of Michigan sMDT prototype chambers for
  the HL-LHC ATLAS muon detector upgrade}},
\newblock \bibinfo{journal}{JINST} \bibinfo{volume}{16} (\bibinfo{year}{2021})
  \bibinfo{pages}{P11027}.
%Type = Article
\bibitem[{Aad et~al.(2019)}]{ATLAS:2019mcj}
\bibinfo{author}{G.~Aad}, et~al. (\bibinfo{collaboration}{ATLAS}),
\newblock \bibinfo{title}{{Resolution of the ATLAS muon spectrometer monitored
  drift tubes in LHC Run 2}},
\newblock \bibinfo{journal}{JINST} \bibinfo{volume}{14} (\bibinfo{year}{2019})
  \bibinfo{pages}{P09011}.

\end{thebibliography}

\end{document}